\documentstyle[12pt,amssymb,epsfig]{article}

\newcommand{\eq}[1]{(\ref{#1})}

\def\e{{\rm e}}

\def\d{\partial}
\def\l{\left(}
\def\r{\right)}

\def\be{\begin{equation}}
\def\ee{\end{equation}}

\begin{document}
\title{%
\begin{flushright}
\normalsize 
ULB-TH-01/24\\
UNIL-IPT-01-14
\end{flushright}
\protect\vspace{5mm}
Neutrino masses with a single generation in the bulk.
}
\author{J.-M.~Fr\`ere$^1$, M.V.~Libanov$^{2}$ and
S.V.~Troitsky$^{2,3}$\\
\small\em
$^{1}$~Service de Physique Th\'{e}orique, CP 225,\\
\small\em
  Universit\'{e} Libre de Bruxelles, B--1050, Brussels, Belgium;\\
\small\em
$^{2}$~Institute for Nuclear Research of the Russian Academy of
Sciences,\\
\small\em
60th October Anniversary Prospect 7a, 117312, Moscow, Russia;\\
\small\em
$^{3}$~Institute of Theoretical Physics, University of Lausanne,\\
\small\em
CH-1015, Lausanne, Switzerland
\small\em
}
\date{}
\maketitle
\vspace{-12mm}
\begin{abstract}
In a class of multidimensional models, topology of the thick brane
provides three chiral fermionic families with hierarchical masses and
mixings in the effective four-dimensional theory, while the full model
contains a single vector-like generation. We discuss how to
incorporate three non-degenerate neutrino masses in these models with
the help of only one singlet bulk fermion.
\end{abstract}
\newpage
\section{Introduction}
One of the interesting possibilities which open up in theories with
more than four spacetime dimensions is to explain the misterious
pattern of fermion mass hierarchies \cite{overlaps,LT,FLT}. In
previous works \cite{LT,FLT}, we have shown how a single family of
fermions, with vectorlike couplings to the Standard Model group $SU(3)
\times SU(2) \times U(1)$ in 6 dimensions was reduced to $n$ {\it
chiral} families in 4 dimensions. This mechanism hinged on a
localisation on a vortex in the "transverse" dimensions, characterized
by a winding number $n$. As a variant, an effective vortex achieving
the same result could be simulated by coupling the fermions to a
winding-number 1 scalar elevated to the $n$th power.

The value $n=3$ is not favoured in such schemes, and was only invoked
for its phenomenological interest. We also showed how a relatively
simple scalar structure (the usual Higgs doublet $H$, supplemented by
the vortex-defining field $\Phi$) suffices to generate the mass
hierarchy of the $n$ families, while an auxiliary field $X$ could be
used to generate a mixing taking place mostly between adjacent
generations.

The key to the construction is that interaction with a vortex
with winding number $n$ leads, by the index theorem \cite{index}, to $n$
chiral
zero modes in four dimensions, and this for each fermion species
coupled to the structure.

We now return to this construction and pay special attention to the
case of neutrinos.  The simplest attitude (which was in a way
implicit in the previous papers) would be to treat neutrinos just like
the other fermions, that is to include both left- and right-handed
fields, say $\nu$  and  $N$ in four dimensions, and to generalize them
to
six dimensions. Dirac masses and mixings are then obtained just like
those of
the charged fermions.  There is nothing fundamental against this
approach, except the very small value of the ratio between neutrino
and charged fermion masses (typically less than $ 10^{-6}$).  This
would require severe fine tuning, either directly at the level of the
Yukawa couplings themselves, or in a more covert fashion through a
more complicated scalar structure (see footnote below).

We therefore try to explore other ways of getting small neutrino
masses.

It has become standard practice \cite{NeutrinoED} 
in the context of extra (flat and compactified)
dimensions, to assume that right-handed neutrions, being gauge
singlets, could propagate in the "bulk" of space, escaping
confinement\footnote{Other possibilities include 
putting left-- and right-handed neutrinos to very different places 
inside the thick brane \cite{LRnuthickbrane} or putting the scalar
(not the fermion) in the bulk. The latter approach can be easily
implemented in the model of Ref.~\cite{FLT} where additional scalar
fields, others than the electroweak Higgs doublet, participate in mass
generation.}. This is of course possible since such particles do not
participate directly in the gauge interactions, and therefore do not
affect (at least at tree level, see for instance Ref.~\cite{Giudice})
the behaviour of electroweak exchanges.  In such a context, the
smallness of the neutrino mass stems not from artficially tuned
Yukawa couplings, but from the severely reduced overlap between the
confined and "bulk" wave functions.  Could such a strategy work here?
The transcription of such a mechanism in our context involves the
introduction of a six-dimensional fermion field, $N$, uncoupled to the
vortex
field $\Phi$.  This however is in general not sufficient, if the
two-dimensional transverse 
space stays unbounded (a point which we did not really need to specify
this far, since the vortex achieved in any case the localisation of
physical fields in a four-dimensional tube). Using $\theta$ and $r$ as
variables to describe the two extra dimensions, we should now require
$r<R$, while keeping $R \gg r_{vortex}$, to avoid perturbing the
previous construction.  Is this step sufficient?  In principle, if
$1/R$ were close to the expected light neutrino mass scale (say a
fraction of an eV), we would have a whole set of Kaluza-Klein towers
at hand, each with relatively light states to couple to the $n$
left-handed neutrinos.  This line is however somewhat dangerous, since
it may lead to a conflict with data on the evolution of supernovae
\cite{SN}.  We will
avoid this situation by requesting $1/R \gg m_{\nu}$. For simplicity we
may even take\footnote{This is to avoid problems with weak
interactions universality. We suppose that the Standard Model gauge
fields are localized inside the brane by means of a mechanism
\cite{DvaliShifman,Akhmedov} which preserves charge universality by
itself, see discussion in Ref.~\cite{RuSergd}.} $1/R \gg 100$~GeV.

This further requirement however leads us to a pecular situation: we
may not have enough light "right-handed" partners to provide masses to
the $n$ neutrinos! Indeed, we typically expect only one fundamental
mode of zero mass.  Introducing $n$ such fermions would of course solve
the problem, but would be in opposition to the approach, where at
least the charged family multiplication simply results from the
topological structure.  Fortunately, even with the simplest use of one
$N$ field, we get easily two light states (this is a funciton of the
chosen charge assignements, see below), leaving only $n-2$ degenerate
massless $\nu_L$ states.  Since one strictly massless neutrino is no
problem (only differences in masses squared are tested this far), the
neutrino degeneracy is thus solved for $n \leq 3$.

The scheme presented below uses one extra scalar field to match
quantum numbers and generate the neutrino masses. It is of course also
possible to induce mixing terms, but these are probably redundant,
since off-diagonal terms are already allowed for in the charged
leptons sector.

\section{Models with a single generation in extra dimensions}
If different fermionic modes have
different wave function profiles in extra dimensions, then their
overlaps with the Higgs wave function may produce hierarchical
structure of masses and mixings \cite{overlaps}. 
In the class of models \cite{LT,FLT}, each
multi-dimensional fermion develops three chiral zero modes localized
on a four-dimensional brane due to topological
properties of the brane background.  The Index theorem guarantees that
the three zero modes are linearly independent, and thus have different
profiles in extra dimensions. Analysis of the equations for these zero
modes demonstrates that a hierarchy in the mass matrix indeed appears
due to overlaps of the wave functions. For the discussion of this
mechanism and
comparison with other approaches, see Ref.~\cite{LT}.

We use conventions of Ref.~\cite{LT} throughout the paper.  We will
work in a specific realisation of this approach which was suggested in
Ref.~\cite{FLT}. The model is formulated in six dimensions, and the
brane is an Abelian vortex made of the gauge field of $U(1)_g$ gauge
group and a scalar field $\Phi$. Field content of the model of
Ref.~\cite{FLT} is given in Table~1 for easy reference 
\begin{table}
\begin{center}
\begin{tabular}{|rc|c|c|c|c|c|}
\hline
\multicolumn{2}{|c|}{fields}
& profiles&\multicolumn{2}{|c|}{charges}&
\multicolumn{2}{|c|}{representations}\\
\cline{4-7}
&&&$U(1)_g$&$U(1)_Y$&$SU(2)_W$&$SU(3)_C$\\
\hline
scalar&$\Phi$&$F(r)\e^{i\theta}$&+1&0&{\bf 1}&{\bf 1}\\
&&$F(0)=0$, $F(\infty)=v_\Phi$&&&&\\
\hline
scalar&$X$&$X(r)$&+1&0&{\bf 1}&{\bf 1}\\
&&$X(0)=v_X$, $X(\infty)=0$&&&&\\
\hline
scalar&$H$&$H(r)$&$-1$&$+1/2$&{\bf 2}&{\bf 1}\\
&&$H(0)=v_H$, $H(\infty)=0$&&&&\\
\hline
fermion&$L$&3 L zero modes&axial $+3/2$&$-1/2$&{\bf 2}&{\bf 1}\\
\hline
fermion&$E$&3 R zero modes&axial $-3/2$&$-1$&{\bf 1}&{\bf 1}\\
\hline
\end{tabular}
\end{center}
\caption{Field content of the model of Ref.~\protect\cite{FLT}
(scalars and leptons only).
}
\label{table:fields}
\end{table}
(we concentrate here on the leptonic sector since we are interested in
neutrino masses). Like other Standard Model fermions, leptons are zero
modes of the six-dimensional Dirac spinors, namely, of the electroweak
$SU(2)$ doublet $L$ and $SU(2)$ singlet $E$
which have axial charges +3/2 and $-3/2$, respectively, under
$U(1)_g$. The interaction with vortex field,
\be	
g_l\Phi^3\bar L{1-\Gamma_7\over 2} L+
g_e\Phi^{*3}\bar E{1-\Gamma_7\over2} E+{\rm h.c.},
\label{n5**}
\ee results in three localized left- (right-)handed in
four-dimensional sence zero modes of $L$ ($E$) which describe three
generations of usual leptons. The zero modes have the form
$$
L_p\sim
\left(
\begin{array}{c}
0\\
{\bf l}_p l_{p}(r){\rm e}^{ip\theta}\\
{\bf l}_p l_{2-p}(r){\rm e}^{-i(2-p)\theta}\\
0
\end{array}
\right), 
~~~~
E_p\sim
\l
\begin{array}{c}
{\bf e}_p e_{2-p}(r)\e^{-i(2-p)\theta} \\
0\\
0\\
{\bf e}_p e_{p}(r)\e^{ip\theta} \\
\end{array}
\r,
$$
where $r$, $\theta$ are polar coordinates in the bulk, $p=0,1,2$
enumerates three zero modes which in the effective four-dimensional
Lagrangian are
described by
two-component Weyl spinors ${\bf l}_p$, ${\bf e}_p$, left- and
right-handed, respectively. The radial
functions $L_p(r)$, $e_p(r)$ have the following leading behaviour:
$$
l_p(r) \sim r^p, r\to 0; ~~~
l_p(r)\sim \e^{-g_l vr}, r\to\infty,
$$
and the similar for $e_p(r)$. 

The scalar potential
$$
{\lambda\over 2}\l |\Phi|^2-v^2\r^2+
{\kappa\over 2} \l |H|^2-\mu^2\r^2+h^2|H|^2|\Phi|^2+
{\rho\over 2} \l |X|^2-v_1^2\r^2+\eta^2|X|^2|\Phi|^2
$$
results in localisation of $H$ and $X$ inside the vortex.
Interaction with the Higgs field,
\be
YHX\bar L\frac{1-\Gamma_7}{2}E+Y\epsilon H\Phi\bar
L\frac{1-\Gamma_7}{2}E+
{\rm h.c.},
\label{n5*}
\ee
where $Y$ and $Y\epsilon$ are two Yukawa couplings, provides a
hierarchical structure of masses and mixings of charged leptons, as is
discussed in Refs.~\cite{LT,FLT}. The reason for hierarchy is the
different behaviour of three zero modes inside the brane which, after
integrating out extra dimensions, results in the hierarchical structure
of overlaps of wave functions for relatively narrow Higgs field (see
the sketch at Fig.~1).
\begin{figure}
\centerline{\epsfxsize=0.9\textwidth \epsfbox{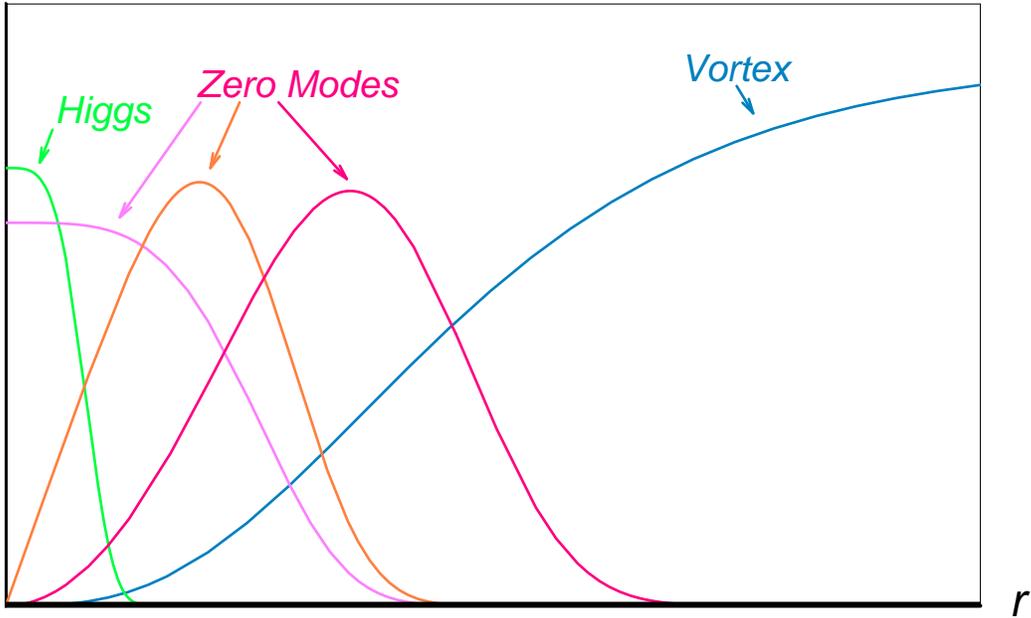}}
\caption{Sketch of the wave function profiles inside the brane.}
\end{figure}

\section{Neutrino masses}
At the next step we follow the usual approach \cite{NeutrinoED} to
neutrino masses in models with extra dimensions, namely, we put a
singlet fermion in the bulk. It plays the role of the usual
right-handed neutrino of the see-saw models. The Dirac masses of the
localized neutrinos are suppressed by the bulk size. The usual
approach is to put three singlets in the bulk, otherwise only one
linear combination of localized modes gets mass. This is in some
contradiction with our intent to have a single generation in the
higher dimensions. However, with more than one extra dimensions the
bulk spinor has, in principle, enough components to give masses to
more than one four-dimensional neutrino in the effective theory
\cite{Agashe}. Here, we exploit this fact in a slight modification of
our model of Ref.~\cite{FLT}.

Let us introduce the six-dimensional Dirac spinor $N$, which is a
Standard Model
singlet and has no half-integer axial charge under $U(1)_g$. In this
case, no
interaction term similar to Eq.~\eq{n5**}, which could result in
localisation of the zero modes of $N$, can be written. The free spinor
$N$ satisfies the six-dimensional Dirac equation,
$$
i\Gamma_A\d_A N=0.
$$
It is convenient to perform the four-dimensional Fourier transform
first,
$$
N=\int d^4k \e^{-ik^\mu x_\mu}\tilde N(K_\mu,r,\theta),
$$
substitute to the Dirac equation and multiply the latter by $\Gamma_0$,
$$
(k_0-k_i\Gamma_0\Gamma_i)\tilde N=D\tilde N,
$$
where the operator $D=i\Gamma_0\Gamma_\alpha\d_\alpha$ anticommutes with
the
operator in the left hand side. The latter fact allows us to look for
a solution expanded in series of eigenvectors of $D$ with eigenvalues
$\omega$,
$$
D\tilde N=\omega\tilde N.
$$
The solution to this equation can be found in Appendix~\ref{App1}. We
note here that the zero mode $\omega=0$ which is regular both in the
origin and at infinity consists of just four linearly independent
constant two-component spinors,
\be
\tilde N=\l
\begin{array}{c}
{\bf n}_1\\
{\bf n}_2\\
{\bf n}_3\\
{\bf n}_4
\end{array}
\r.
\label{n6*}
\ee From the four-dimensional point of view, we have two left-handed
Weyl spinors ${\bf n}_{2,3}$ and two right-handed Weyl spinors ${\bf
n}_{1,4}$ (see discussion of chirality in six and four dimensions in
ref.~\cite{LT}). The latter two independent spinors, indeed, can be
used to give masses to two of three localized neutrinos. This provides
three $\Delta m^2$ which appear in the usual four-dimensional neutrino
models.

The advantage of having an extra Weyl spinor, however, disappears if
we use interactions with $H$, $X$, and $\Phi$, similar to
Eq.\eq{n5*}. Indeed, $U(1)_g$ invariance requires that $N$ has the
charge $3/2$ under this group (half-integer {\em axial} charges are not
allowed in order not to trap the $N$ particle on the brane) for the term
like
$$
HX\bar L {1-\Gamma_7\over 2}N
$$
(or $-3/2$ if we use $(1+\Gamma_7)/2$). Without the projector $(1 \pm
\Gamma_7)/2$, the term is not gauge invariant; on the other hand, this
projector kills one of two spinors ${\bf n}_{1,4}$.  To overcome this
difficulty, we introduce a new scalar field $S$ which is analogous to
$X$ but has a charge $1/2$ under $U(1)_g$ (the two fields, $N$ and
$S$, which we introduce in this paper, are described in Table
2\footnote{With these charge assignments, Dirac mass term for $N$ is
not forbidden. The simplest way to forbid it is to assign a mixed,
axial and vector, charge to $N$. The axial charge $q_A$ should not be
half-integer, otherwise $N$ is localised on the brane. Together with
the charge $q_S$ of $S$, they satisfy $q_S=3/2+q_A$, while the vector
charge of $N$, $q_V=1$. For example, one of the solutions is
$q_A=1/4$, $q_V=1$, $q_S=7/4$.}).
\begin{table}
\begin{center}
\begin{tabular}{|rc|c|c|c|c|c|}
\hline
\multicolumn{2}{|c|}{fields}
& profiles&\multicolumn{2}{|c|}{charges}&
\multicolumn{2}{|c|}{representations}\\
\cline{4-7}
&&&$U(1)_g$&$U(1)_Y$&$SU(2)_W$&$SU(3)_C$\\
\hline
fermion&$N$&bulk modes&$+1$&0&{\bf 1}&{\bf 1}\\
\hline
scalar&$S$&$S(r)$&$+3/2$&0&{\bf 1}&{\bf 1}\\
&&$S(0)=v_S$, $S(\infty)=0$&&&&\\
\hline
\end{tabular}
\end{center}
\caption{Fields 
introduced to give masses to the neutrinos.
}
\label{table:newfields}
\end{table}
With these charge assignements, we can write down two interaction
terms,
\be
y_1 HS\bar L{1-\Gamma_7\over 2}N,
\label{n7*}
\ee
and
\be
y_2 HS^*\bar L{1+\Gamma_7\over 2}N.
\label{n7**}
\ee
Substituting zero modes of $l$ and $N$, one gets the effective
Lagrangian
$$
\int\! r dr d\theta \,
\left(y_1 HS(r) l_{2-p}(r){\rm e}^{-i(2-p)\theta}{\bf l}_p
{\bf n}_1+ y_2
HS(r) l_{p}(r){\rm e}^{ip\theta}{\bf l}_p {\bf n}_4
\right)
$$
Since $H$ and $S$ have no $\theta$ dependence in their classical
profiles, the integral over $\theta$ is non-zero for $p=2$ (the first
term) and $p=0$ (the second). Thus, the modes ${\bf l}_{0,2}$ and
${\bf n}_{4,1}$ are paired in the mass matrix. The masses are given by
integrals
$$
y_{1,2} 2\pi\int\! r dr\, HS(r) l_{0} (r) |n_{4,1}|
$$
where the norm $|n_i|\sim 1/R$, $R$ being the size of the extra
dimensions. Masses of ${\bf l}_0$ and ${\bf l}_2$ differ only by
Yukawa couplings $y_{1,2}$ in two terms.

Two notes are in order. First, would we add another scalar with the
same quantum numbers as $S$ but with winding number 1, we would receive
a
more complicated mass matrix (mixing would appear in the neutrino
sector as well as in the sector of charged leptons), but not the mass
for the third neutrino -- because we have only two right-handed Weyl
spinors, ${\bf n}_1$ and ${\bf n}_4$, available. One linear
combination of ${\bf l}_p$ would still be massless. Second, the
Eqs.~\eq{n7*}, \eq{n7**} mix ${\bf l}_p$ with nonzero modes of $N$ as
well. However,
to satisfy simultaneously constraints from weak charge universality
and from Supernova 1987a, these mixings should be negligible.

\section{Parameters and estimates}
In order to count parameters and to estimate the scales involved in
the model, we make use of the ``step Higgs'' approximation (full
numerical analysis will be presented elsewhere \cite{Emin}). First of
all, recall that in six dimensions, a scalar field has dimension
(mass)$^2$ and a fermion field has dimension (mass)$^{5/2}$. This means
that the transverse part of the fermion, described by $l_p(r)$, has
dimension (mass)$^1$ (four-dimensional spinors, denoted by letters in
bold face, are of course (mass)$^{3/2}$). Dimensions of coupling
constants are $[Y]={\rm (mass)}^{-3}$, 
$[g]={\rm (mass)}^{-5}$, 
$[h]={\rm (mass)}^{-1}$,
$[\lambda]={\rm (mass)}^{-2}$. The fundamental dimensionful parameters
of
the theory are the vortex scale $v$ (in theories where large extra
dimensions are invoked to reformulate the gauge hierarchy problem, $v$
is of order multi-dimensional Planck mass), and the compactification
scale $R$. It is convenient to rescale the parameters of the theory:
$$
\tilde g_{l,e}=g_{l,e} v^{5/2},~
\tilde v = v^{1/2},~
\tilde \mu=\mu v^{-1/2},~
\tilde v_1=v_1 v^{-1/2},~
\tilde h=h v^{1/2},~
\tilde Y=Y v^{3/2},~
\tilde\lambda=\lambda v.
$$
Tilded coupling constants $\tilde g$, $\tilde h$, $\tilde Y$,
$\tilde\lambda$  are
dimensionless, while $\tilde v$, $\tilde v_1$, $\tilde \mu$ have
dimension (mass)$^1$.

In the narrow Higgs approximation, $H(r)$ can be substituted by
$$
H(r)\simeq\left\{
\begin{array}{cl}
H(0),& r\le r_H;\\
0, & r>r_H,
\end{array}
\right.
$$
where $r_H\sim(\tilde h \tilde v)^{-1}$ is the width of $H(r)$ which
is supposed to be smaller than widths of other profiles. Then, we can
use the following approximations for $r<r_H$:
$$
X(r)\approx X(0),
~~~
S(r)\approx S(0),
$$
$$
l_p(r)\approx \sqrt{\tilde g_l}\tilde v \left(\sqrt{\tilde g_l}\tilde
v r \right)^p,
$$
$$
e_p(r)\approx \sqrt{\tilde g_e}\tilde v \left(\sqrt{\tilde g_e}\tilde
v r \right)^p,
$$
The coefficients at $r^p$ in fermionic radial functions are
determined, approximately, in Ref.~\cite{Emin}. For the scalar fields
$H$, $X$, and $S$, the boundary conditions at $r=0$ are $\d H(r)/\d
r=0$, etc.; so that $H(0)$, $X(0)$, $S(0)$ are to be determined by
solving the full nonlinear system \cite{Emin}. We denote $H(0)=v_H$,
$X(0)=xv_H$, $S(0)=sv_H$, where $x$ and $s$ are coefficients of order
one, $v_H$ depends non-trivially on $\mu$ and $v$. Since $\mu$ and $v$
are independent parameters of the lagrangian, $v_H$ does not need to
be very close to $v$, though we do not wish them to differ too much in 
order to avoid fine tuning of the parameters.

Diagonal elements of the mass matrix of charged leptons are, to this
approximation,
$$
m_{pp}\approx 2\pi Y H(0) X(0)\int_0^{r_H}\!r\,dr\,l_p(r) e_p(r)
\approx \pi\tilde Y x {\tilde v^2_H\over \tilde v}{\delta^{p+1}\over
p+1},
$$
where $\delta=\sqrt{\tilde g_l\tilde g_e}/\tilde h^2$ should be
sufficiently small to provide the mass hierarchy between three
generations. Smallness of $\delta$ exactly corresponds to the fact
that $H(r)$ is more narrow than $l(r)$, $e(r)$.

Non-diagonal mass matrix elements are estimated in a similar way,
taking into account the behaviour of $F(r)$ at small $r$:
$$
F(r)\sim v \left(\sqrt{\tilde\lambda}vr\right).
$$
The result is that
$$
m_{p,p-1}\approx m_{pp}\tilde\epsilon,
$$
where generation-independent constant
$$
\tilde\epsilon={\epsilon\over x}{\tilde v\over\tilde
v_H}\sqrt{\tilde\lambda \over \tilde g_e}.
$$
This corresponds to the mass matrix of charged leptons
\be
M_e\approx M_{0e}
\left(
\begin{array}{ccc}
1&0&0\\
\tilde\epsilon&\delta/2&0\\
0&\tilde\epsilon\delta/2&\delta^2/3
\end{array}
\right)
,
\label{4.3*}
\ee
where the overall constant $M_{0e}=\pi x\tilde Y\tilde v_H^2/\tilde
v$.

It is straightforward to obtain, to the same approximation, the
neutrino mass matrix
$$
M_\nu=M_{0\nu}
\left(
\begin{array}{cc}
1&0\\
0&0\\
0&\tilde y_2/\tilde y_1
\end{array}
\right)
,
$$
where 
$$
M_{0\nu}=\sqrt{\pi}\tilde y_1{\tilde g_l\over \tilde h^2}{\tilde
v^2_H\over\tilde v}\,{1\over\tilde v R}
$$
and we rescaled, as before, $\tilde y_{1,2}=y_{1,2} v^{3/2}$. The
matrix is $3\times 2$ because we have only two light right-handed
modes, ${\bf n}_1$ and ${\bf n}_4$.

Thus, up to dimensionless constants of order one, the neutrino mass
scale is suppressed by a factor of $\displaystyle{1\over\tilde v R}$
with respect to the mass scale of charged fermions, $\tilde
v^2_H/\tilde v$. One can have simultaneously $\tilde
v^2_H/\tilde v\sim 100$~GeV, $m_\nu\sim 0.1$~eV, and Planck mass
relation $\tilde v^2R\sim 10^{19}$~GeV satisfied for $\tilde v\sim
1000$~TeV, $\tilde v_H\sim 10$~TeV, $R\sim 10^{-6}$~mm.

In our model, neutrino mass matrix does not contain mixings. In the
sector of charged leptons, however, mixing appears, see
Eq.~\eq{4.3*}. Three parameters $M_{0e}$, $\delta$, and
$\tilde\epsilon$, determine the masses of three charged leptons; one
can fit known values of $M_{e,\mu,\tau}$ with $M_{0e}\approx 650$~MeV,
$\delta\approx 0.13$, $\tilde\epsilon\approx 2.5$. The latter value
means
that mixing is not negligible in the leptonic sector. Since all
leptonic mixings appear in Eq.~\eq{4.3*}, our model {\bf predicts}
values of the neutrino mixing angles which are implicitly determined
by masses of charged leptons. In usual notations, they correspond
to
$\sin\theta_{12}\sim 0.14$, $\sin\theta_{13}\sim 0.37$,
$\sin\theta_{23}\sim 0.99$.
On the other hand, values of two $\Delta m^2$ are predicted only
by the order of magnitude since they depend on unknown constants
$\tilde y_{1,2}$.

\section{Conclusions}
In a class of multidimensional models with one vector-like fermionic
family, the low-energy effective theory describes three chiral
families in four dimensions. Hierarchy of fermionic masses appears
without fine tuning of parameters. In this paper, we have shown that
nondegenerate neutrino masses can be incorporated in a model of this
class with the help of only one multidimensional Dirac fermion. One of
three neutrinos is exactly massless\footnote{See, however,
Appendix~\ref{App3}} while masses of two others are parametrically
close to each other. Neutrino mass matrix is diagonal, but mixings in
the leptonic sector are present in the mass matrix of charged
leptons. Once masses of $e$, $\mu$, $\tau$ are fixed to their
experimental values, the model {\em predicts} neutrino mixing angles,
$\sin\theta_{12}\sim 0.54$, $\sin\theta_{13}\sim 0.31$,
$\sin\theta_{23}\sim 0.99$. Of course, these values were obtained with
very rough approximations, and should not be taken too
seriously. Neutrino mass differences are predicted by order of
magnitude only, $\Delta m^2\sim (0.1~eV)^2$ for natural values of
parameters.

\tolerance=800

The authors are indebted to K.Agashe, E.Akhmedov, T.Gherghetta,
M.Giovannini,
I.Gogoladze, A.Neronov, E.Nugaev, V.Rubakov, G.Senjanovic,
M.Shaposhnikov, S.Sibiryakov, A.Smirnov for helpful discussions on
different subjects related to this work.  The work of M.L.\ and S.T.\
is supported in part by RFFI grant 99-02-18410a, by the Council of
Presidential Grants and State Support of Leading Scientific Schools,
grant 00-15-96626, by CRDF award RP1-2103, and by the programme SCOPES
of the Swiss National Science Foundation, project No.~7SUPJ062239,
financed by Federal Department of Foreign affairs.  The work of S.T.\
is supported in part by Swiss Science Foundation, grant 21-58947.99.
This work was initiated during the visits of M.L.\ and S.T.\ to ULB,
which we thank for kind hospitality and partial support by the
``Actions de Recherche Concret\'ees'' of ``Communaut\'e Fran\c{c}aise
de Belgique'' and IISN--Belgium. S.T.\ thanks High Energy Group of
ICTP (Trieste) for hospitality during his visit, when part of this
work was done and numerous discussions were carried out.

\appendix
\section{Solution to the Dirac equation for $N$}
\label{App1}

In this Appendix, we present the solution to the transverse part of
Dirac equation,
$$
D\tilde N=\omega \tilde N,
$$
where $D=i\Gamma_0\Gamma_\alpha\d_\alpha$.
The general solution for $\omega=0$ is
$$
\tilde N=\l
\begin{array}{c}
{1\over {\cal{N}}_1} {\bf n}_1 r^{-p_1}\e^{ip_1\theta}\\
{1\over {\cal{N}}_2} {\bf n}_2 r^{p_2}\e^{ip_2\theta}\\
{1\over {\cal{N}}_3} {\bf n}_3 r^{-p_3}\e^{ip_3\theta}\\
{1\over {\cal{N}}_4} {\bf n}_4 r^{p_4}\e^{ip_4\theta}
\end{array}
\r.
$$
Here, ${\bf n}_a$ are independent two-component spinors. The solutions
regular both at small and large $r$ correspond to $p_a=0$,
Eq.~(\ref{n6*}). 
For them, the normalization factors ${\cal{N}}_a=\sqrt{\pi}R$,
where $R$ is the radius of the bulk.

For $\omega\ne 0$, the regular at $r=0$ solution depends on two
two-component spinors ${\bf n}$, ${\bf m}$ for each $\omega$:
\be
\tilde N=\l
\begin{array}{c}
{\bf n}_\omega J_p(\omega r)\e^{ip\theta}\\
-{\bf n}_\omega J_{p-1}(\omega r)\e^{i(p-1)\theta}\\
{\bf m}_\omega J_q(\omega r)\e^{iq\theta}\\
{\bf m}_\omega J_{q-1}(\omega r)\e^{i(q-1)\theta}
\end{array}
\r.
\label{zero_modes_of_N}
\ee
A series of eigenvalues $\omega_n^{(p)}$ are determined by the
boundary conditions at large $r$. The exact eigenvalues depend on the
compactification scheme chosen (see discussion in Appendix \ref{App3}),
but in
general, they are of order
$$
\omega_n^{(p)}\sim\pi n/R.
$$

\section{Compactification}
\label{App3}

The compactification scheme defines boundary conditions at large
$r$. It should satisfy the following criteria:

(i) it should preserve three zero modes with different winding numbers
of the fields localized on the vortex;

(ii) it should allow for at least one (Dirac) zero mode of the bulk
singlet.

The second requirement is essentially equivalent to compactification
on a Ricci-flat manifold (that is, a manifold with zero scalar
curvature ${\cal{R}}$) \cite{Witten}. Toroidal compactifications satisfy
this
requirement since the torus have ${\cal{R}}=0$. However, the
corresponding boundary conditions usually
kill one or two of three zero modes because they allow either even or
odd winding numbers for a given field. Compactification of a vortex on
a sphere preserves three zero modes \cite{Nakamura} but does not
satisfy the second requirement: for a sphere, ${\cal{R}}\ne 0$. This
would open an interesting possibility of having three massless
neutrinos and explaining observed neutrino anomalies by mixing with
Kaluza-Klein modes only (this would lead to a nonconventional pattern
of oscillations, as was discussed in Ref.~\cite{Brussels} in the
frameworks of a toy model). However, this possibility is in
contradiction with bounds on mixing with Kaluza-Klein states from
supernova evolution.

One can also keep the compactification mechanism unspecified, and
consider flat space of finite volume. This means that the fields will
live inside a cilinder of large radius $R$ in six dimensions. The
corresponding boundary condition, that is, absence of fermionic
current through the boundary, can be reformulated \cite{bags} as
$$
i\Gamma_A\d_A\psi=\psi
$$
for a Dirac spinor $\psi$. Applied to $N$, Eq.~\eq{zero_modes_of_N},
this boundary condition selects not only constant, but also rather
peculiar zero modes which grow up at large $r$. In this case, it is
not necessary to introduce the additional field $S$ to obtain a second
neutrino mass, and it suffices to use the additional mode,
alternatively, with the field $S$ included, 3 (even 4 if the model is
extended) neutrino masses can be generated, although the last 2 are
then hierarchically smaller than the ones previously defined.  At this
level however, such modes appear more as a quirk of a particular way
to implement the finite volume.

It seems the most promising to have a finite volume in the case of
warped
transverse dimensions. Particularly interesting is the fact that there
exist six-dimensional solutions with gravity localized on an
Abelian vortex, Ref.~\cite{WarpedVortex}. These six-dimensional
solutions provide also a natural framework for the localization of gauge
fields at weak coupling, thus opening the way to fully realistic and
calculable models.

\end{document}